\documentclass[conference]{IEEEtran}
\IEEEoverridecommandlockouts
\usepackage{cite}
\usepackage{amsmath,amssymb,amsfonts}
\usepackage{algorithmic}
\usepackage{graphicx}
\usepackage{textcomp}
\usepackage{xcolor}
\usepackage[switch]{lineno}
\usepackage{url,hyperref}
\usepackage{adjustbox}
\usepackage{float}
\usepackage[T1]{fontenc}
\usepackage[utf8]{inputenc}
\usepackage{tikz}
\usepackage{pifont}
\usepackage{scalerel}
\usepackage{ifthen}

\def\BibTeX{{\rm B\kern-.05em{\sc i\kern-.025em b}\kern-.08em
    T\kern-.1667em\lower.7ex\hbox{E}\kern-.125emX}}
\begin{document}

\title{Towards a privacy-preserving distributed cloud service for preprocessing very large medical images}

\author{Yuandou Wang$^{1}$, Neel Kanwal$^{2}$, Kjersti Engan$^{2}$, Chunming Rong$^{2}$, Zhiming Zhao$^{1,3}$\\
$^1$Multiscale Networked Systems, University of Amsterdam, the Netherlands\\
$^2$Department of Electrical Engineering and Computer Science, University of Stavanger, Norway\\
$^3$LifeWatch ERIC Virtual Lab and Innovation Center (VLIC), Amsterdam, the Netherlands\\
Email: \{y.wang8, z.zhao\}@uva.nl; \{neel.kanwal, kjersti.engan, chunming.rong\}@uis.no\\
}

\maketitle
\begin{abstract}
Digitized histopathology glass slides, known as Whole Slide Images (WSIs), are often several gigapixels large and contain sensitive metadata information, which makes distributed processing unfeasible. Moreover, artifacts in WSIs may result in unreliable predictions when directly applied by Deep Learning (DL) algorithms. Therefore, preprocessing WSIs is beneficial, e.g., eliminating privacy-sensitive information, splitting a gigapixel medical image into tiles, and removing the diagnostically irrelevant areas. This work proposes a cloud service to parallelize the preprocessing pipeline for large medical images. The data and model parallelization will not only boost the end-to-end processing efficiency for histological tasks but also secure the reconstruction of WSI by randomly distributing tiles across processing nodes. Furthermore, the initial steps of the pipeline will be integrated into the Jupyter-based Virtual Research Environment (VRE) to enable image owners to configure and automate the execution process based on resource allocation.  

\end{abstract}


\vspace{-0.75em}
\section{Introduction}
Deep Learning (DL) approaches have advanced and innovated automatic diagnostics, such as quantifying the presence of cancerous cells in digitized histopathology glass slides, called Whole Slide Images (WSIs). However, running these diagnostic services over a large scale requires a significant infrastructure capacity for storing and processing images using complex DL models, e.g., cloud computing and High-Performance computing (HPC), are often needed. The owners of the medical images, e.g., hospitals, often do not have such an infrastructure and have to rely on collaborators with remote infrastructure resources. 

Establishing a DL-based pipeline for medical images on a remote infrastructure is challenging; for instance, 1) WSIs often contain privacy-sensitive information in their metadata and cannot be directly sent out to the public cloud from the hospitals; 2) WSIs are often very large and require high network bandwidth to upload; and 3) WSIs are split into tiles to process~\cite{kanwal2022devil,neel2023} and require specialized hardware, e.g., GPUs, to run complex DL models. Furthermore, it is often complicated to deploy an end-to-end pipeline and create an efficient re-configurable workflow\cite{zhao_special_2009} on remote infrastructure.

\begin{figure}[!htp]
    % \vspace*{-5mm}
    \centering
    \includegraphics[width=0.45\textwidth]{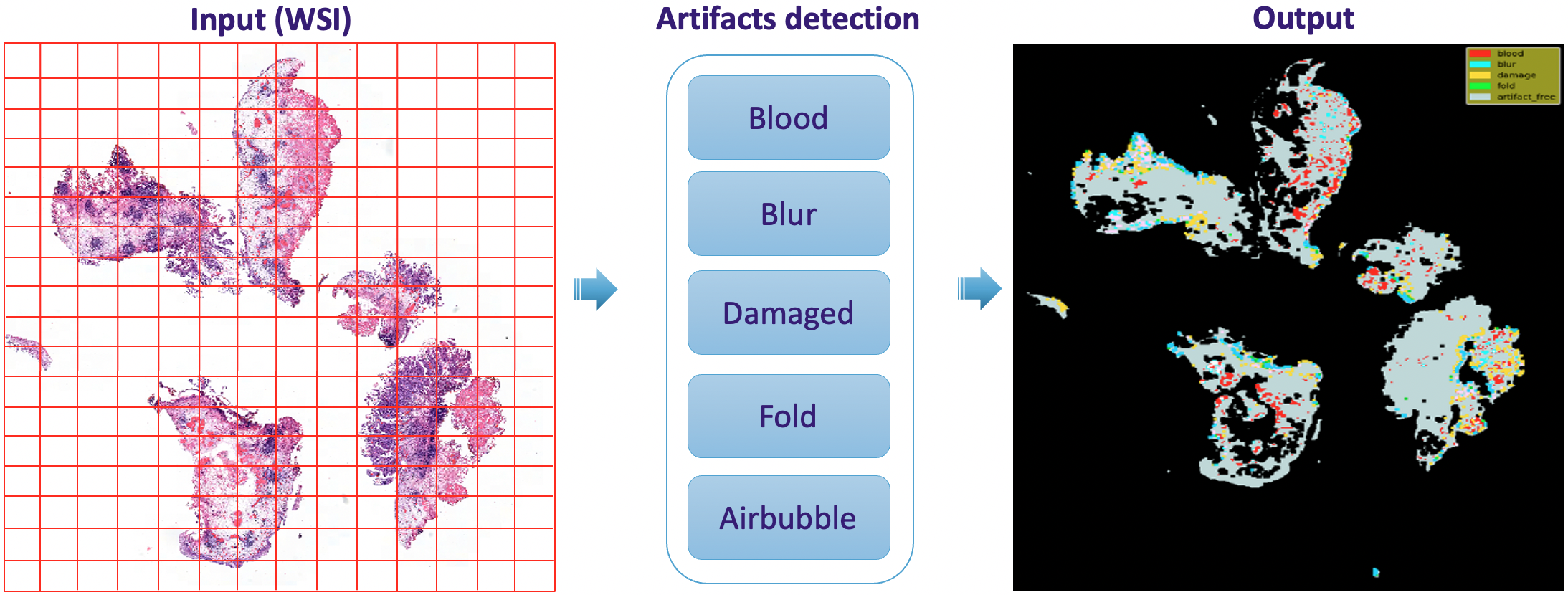}
    \vspace*{-3mm}
    \caption{An overview of the preprocessing pipeline for detecting artifacts in whole slide images.}
    \label{fig: classic WSI preprocessing}
    \vspace*{-6mm}
\end{figure}

In this paper, we will tackle these challenges by proposing a cloud-based service that will be integrated into a collaborative virtual research environment based on the works~\cite{wang2022scaling, zhao2022notebook}, and we will present a use case of a medical image preprocessing application from digital pathology domain to testify our methodology. 

\begin{figure*}[ht!]
    \vspace*{-5mm}
    \centering
    \includegraphics[width=0.8\textwidth]{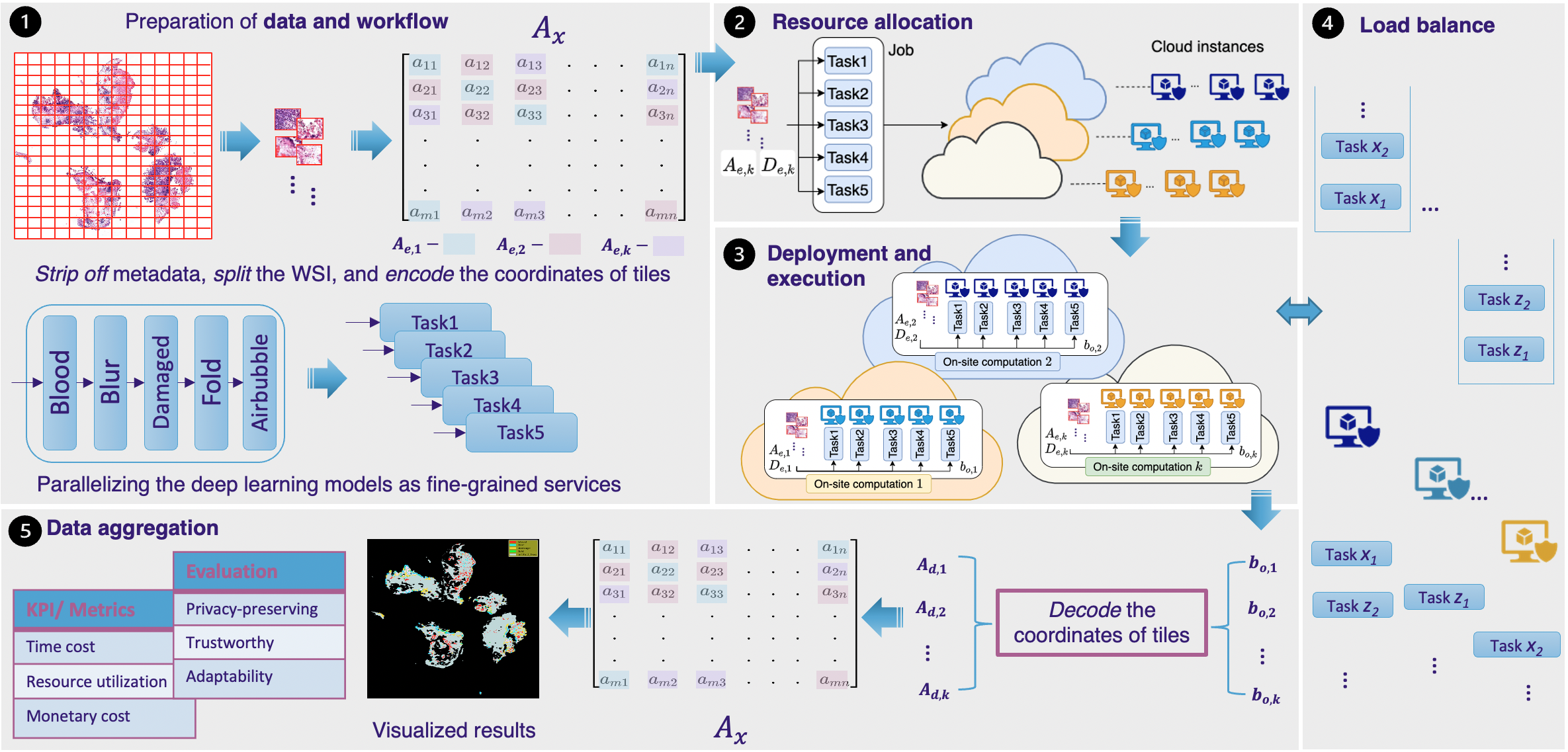}
    \vspace*{-2mm}
    \caption{The methodology proposed in this work. Step~\ding{182} - preparation of data and workflow; step~\ding{183} - resource allocation; step~\ding{184} - deployment and execution; step~\ding{185} - load balance; and step~\ding{186} - aggregating distributed data in a summary view to visualize the final results.}
    \label{fig: scenario}
    \vspace*{-5mm}
\end{figure*}  

\vspace{-0.75em}
\section{Case Study}\label{s:ps}
Digital pathology overcomes the hurdles of traditional histopathology by facilitating the diagnostic process using a WSI~\cite{kanwal2022devil}. The preparation of histological glass slides may result in the appearance of artifacts on the obtained WSI due to improper handling of the tissue specimen during the tissue processing stages. These histological artifacts are diagnostically irrelevant and are usually ignored by pathologists in the diagnosis process~\cite{kanwal2022quantifying, neel2023}. Therefore, it is vital to detect and remove them before applying diagnostic or prognostic algorithms.
Some frequently appearing artifacts are damaged tissue, folded tissue, blur, air bubbles, and diagnostically irrelevant blood~\cite{kanwal2022quantifying,neel2023}. Computational pathology (CPATH) researchers may run DL-based artifact preprocessing algorithms over thousands of WSIs before applying diagnostic algorithms, requiring powerful computational resources to process WSIs efficiently. Fig.~\ref{fig: classic WSI preprocessing} presents an overview of such artifact preprocessing pipeline, which is an ensemble of five DL models for blood, blur, damaged tissue, folded tissue, and air bubbles detection tasks from WSI in a binary fashion. 

Traditionally, the artifact preprocessing pipeline runs over a single machine, bringing the disadvantages of a single security breach or system failure. Besides, handling gigapixel WSIs is time-consuming and resource-intensive, which raises the demand for parallel distributed computing. Nevertheless, WSIs processed on private clouds in research environments are de-identified or pseudonymized under various regulations. This raises concerns about the embedded privacy-sensitive metadata while making distributed processing over public clouds.

\section{Methodology}
We introduce a methodology to cope with the highlighted challenges, as shown in Fig.~\ref{fig: scenario}.
It consists of five main steps: preparation of data and workflow, resource allocation, deployment and execution, load balance, and data aggregation.
% \vspace{-0.75em}
\subsection{Data and workflow preparation}
Step~\ding{182} aims to introduce parallelism and encryption available for the next steps. To guarantee privacy-preserving requirements, we remove the \emph{metadata} from WSI before splitting the gigapixel image into many image tiles to introduce data parallelism. Meanwhile, containerizing the computational tasks as several reusable fine-grained services can improve scalability and security since they are isolated from each other and from the host system. Besides, we apply a matrix $A_x$ to record the distribution of the tiles over a grid which can be encoded to hide its coordinates and divided into sub-matrices $A_{e,1}, A_{e,2}, ..., A_{e,K}$. Each sub-matrix is considered as an index of distributed dataset $D_{e,k}$ for each service-based task. 

\subsection{Resource allocation}
Based on the prepared data and service-based tasks, step~\ding{183} is to map available resources (e.g., clusters at universities and commercial clouds) to various tasks in a manner that optimizes their utilization and satisfies requirements~\cite{hurwicz1973design}. Related methods, such as IC-PCP~\cite{taal2019profiling} and machine learning-based approach~\cite{wang2019multi}, can be improved for workflow scheduling. For bi-objective optimization, such as reducing execution time and cost, there are trade-offs with time performance and monetary cost over the cloud. On this basis, this paper looks into workflow scheduling problems under the influence of privacy requirements and the split data sets, so the research problem is more challenging.  

% \vspace{-0.75em}
\subsection{Deployment and execution}
After a deployment plan is created at step~\ding{183}, the datasets and service-based tasks will be assigned to planned infrastructures equipped with computation, communication, and storage resources. The system should ensure that data storage and task execution remain in place and continue to be effective even among changes (such as downtime, errors, or attackers) to the system or emerging threats, according to step~\ding{184}. Due to distributed processing, it reduces the burden of a single machine and avoids a single security breach. 

\subsection{Load Balance} 
Considering that computing nodes may unpredictably slow down or fail during their execution, step~\ding{185} aims to improve the performance, reliability, and load balance of task-based applications~\cite{soljanin2022technical}. This approach asymptotically achieves near-ideal load balancing and computation cost in the presence of slow nodes (stragglers), which could also be complementary to workflow scheduling.

\subsection{Data aggregation}
Step~\ding{186} takes the predicted distributed output, $b_{o,k}$ for each encoded tile, from step~\ding{184} for every service task, to reconstruct the encoded distributions as $A_{d,k}$.
% Step~\ding{186} first decodes the encrypted distributions $A_{e,k}$ as $A_{d,k}$ fed with the distributed output $b_{o,k}$ generated in step~\ding{184}.
Privacy preservation can be guaranteed by the random-value perturbation technique~\cite{kargupta2005random}. This approach has solid theoretical foundations and is easier to apply for the reconstruction of the encrypted data than others (e.g.,  differential privacy and secure multiparty computation~\cite{li2021privacy}) especially considering a data matrix manipulation.
Then by tracing back coordinates, we can create a segmentation mask for detected artifacts.
It presents the results of the DL-based artifacts detection in a summary view, incl. visualization, evaluation, and metrics. 

\section{System Model}\label{sec: system}
\begin{figure}[H]
\vspace{-2mm}
    \centering
    \includegraphics[width=0.48\textwidth]{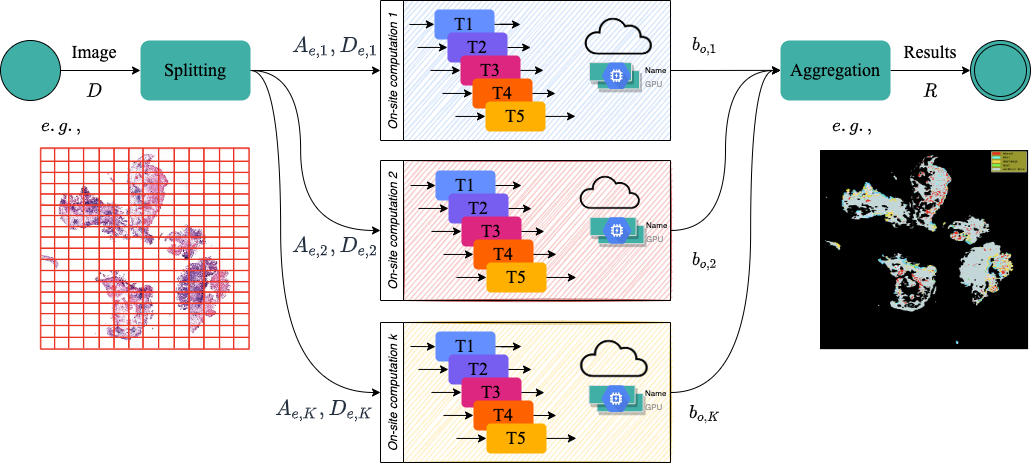}
    \vspace*{-2mm}
    \caption{An overview of the privacy-preserving preprocessing pipeline for whole slide image in a distributed processing manner.
    % It is composed of \textit{Splitting}, \textit{Computation}, and \textit{Aggregation}. Each on-site computation has a set of $\mathcal{N}_T$ tasks to process with accelerated computing resources. Both \textit{Splitting} and \textit{Aggregation} stages are executed on a \textit{Trusted Server}.
    }
    \label{fig: distributed WSI preprocessing}
\vspace*{-3mm}
\end{figure}

% Compared to conventional data processing, this paper proposes to introduce privacy-preserving guarantees, distributed processing, and bi-objective optimization after our technology foresight. 
We sketch out a privacy-preserving distributed processing pipeline for the medical application, shown in Fig.~\ref{fig: distributed WSI preprocessing}. It is composed of three main steps - viz \emph{Splitting}, \emph{Computation}, and \emph{Aggregation}. Both Splitting (See in step~\ding{182}) and Aggregation (See in step~\ding{186}) will be executed on a \emph{Trusted Server}. 
% Meanwhile, suitable data processing requires intensive resources for computation, communication, and storage. 
Such distributed data processing application can be defined as a tuple:
% \begin{footnotesize}
\begin{equation}
    % \mathcal{A}=(\mathcal{M}, \varepsilon, m_{src}, m_{snk}, src, snk,\mathcal{I}, req)
    \mathcal{A}=(\mathcal{M}, \varepsilon, \mathcal{D}, \mathcal{R},\mathcal{I}, req)   
\end{equation}
% \end{footnotesize}
where $\mathcal{M}$ denotes a set of lightweight interconnected \emph{microservices}. A \emph{source microservice} $m_{src}$ processing the data stream produced by the source dataset $\mathcal{D}$. A \emph{sink microservice} $m_{snk}$ representing its final results $\mathcal{R}$. $\varepsilon$ indicates a set of \emph{data streams} $d_{u,i}$ flowing from an upstream microservice $m_u$ to a downstream microservice $m_i\in \mathcal{M}$. $\mathcal{I}$ denotes a set of \emph{cloud infrastructure} and $req$ is a set of user requirements. 

Along with these lines, the research problem has turned the emphasis of studying on privacy-preserving service orchestration -- Or more precisely, \textit{how to customize a virtual infrastructure and schedule the workflow execution under privacy-preserving constraints while reducing its time and monetary cost?}

\noindent \textbf{Privacy requirements:} Reconstruction of a WSI from distributed resource nodes can lead to finding similar medical images using content-based image retrieval and extrapolating possible patient information from other sources. Therefore, using this distributed scheme, their privacy will be preserved during the process. 

\vspace{0.3em}
\noindent \textbf{Bi-objective optimization:} 
Reducing the execution time of the application over the cloud can be crucial for many stakeholders as it can lead to significant cost savings and improve overall processing time for WSI. We aim to reduce the monetary cost $f_1$ and minimize the application's maximum completion time (i.e., makespan) $f_2$. Let us denote $ET(m_{src})$ and $ET(m_{snk})$ as the execution time of splitting and aggregation services over the trusted server. $m_{det}$ for the artifact detection microservices, which will be deployed to the cloud where $ET(m_{det}(A_{e,k}, \mathcal{D}_{e,k}), \mathcal{I}_k)$ denotes its total execution time. Then, the bi-objective optimization problem can be formulated as follows:
\vspace{-0.5em}
\begin{small}
    \begin{align}
        \min f_1 & = \sum_{k=1}^{K} ET(m_{det}(A_{e,k}, \mathcal{D}_{e,k}), \mathcal{I}_k) \times p_k \times x_k \\
        \min f_2 & = ET(m_{src} (\mathcal{D}))+makespan(m_{det}) +ET(m_{snk}(\mathcal{R}))
        % \\vspace{1ex} 
    \end{align}
\end{small}
Here $K$ and $p_k$ represent the number of the split data sets and the unit price of cloud infrastructure $\mathcal{I}_k$, subject to, 
\begin{equation*}
    \begin{small}
        \begin{gathered}
            makespan(m_{det}) = \max\{ET(m_{det}(A_{e,k}, \mathcal{D}_{e,k}), \mathcal{I}_k)\times x_k\}\\
            % \mathcal{M} = \{m_i|0\le i \le \mathcal{N_M}\}, \quad 
            ET(m_{det}(A_{e,k}, \mathcal{D}_{e,k}), \mathcal{I}_k) >0 \\
            % \varepsilon = \{(m_u, m_i, d_{u,i})|(m_u,m_i) \in \mathcal{M} \times \mathcal{M}\}, \\
            % \mathcal{I} = \{\mathcal{I}_j|0\le j \le \mathcal{N_I}\}, 
            % \quad \mathcal{I}_k = \{r_i |0\le i \le \mathcal{N_R}\} \\
          \texttt{where} \hspace{1em}  x_k= \begin{cases}
                1, & \text{if } m_{det} \text{ is mapping to }\mathcal{I}_k,\\
    0,              & \text{otherwise}.
                \end{cases} 
            % R
        \end{gathered}
    \end{small}
\end{equation*}

% where $x_k$ is a boolean value and 1 when $m_{det}$ is mapping to $\mathcal{I}_k$; otherwise 0. 

\section{Discussion and Future Work}
This work-in-progress paper presents the methodology for privacy-preserving task-based parallel applications for distributed cloud environments. Our method enables domain-specific users to handle gigapixel medical images efficiently, maintaining privacy among distributed nodes. In future work, we will develop prototypes and demonstrate the benefits of our pipeline using datasets from different hospitals and integrating the method with a Jupyter-based virtual research environment.   

\section*{Acknowledgment}
This work has been funded by the European Union project CLARIFY (860627), $\text{ENVRI}^\text{FAIR}$ (824068), BlueCloud-2026 (101094227) and LifeWatch ERIC. 

\bibliographystyle{IEEEtran}
\bibliography{main}

\end{document}